\documentclass[aps,prl,twocolumn,showpacs]{revtex4}
\usepackage[english]{babel}
\usepackage{dcolumn}
\usepackage{graphicx}
\usepackage{amsmath}
\usepackage{longtable}
\usepackage{bigstrut}
\usepackage{float}
\usepackage{sidecap}

\begin{document}

\newcommand{\nd}{ND$_3$ }
\newcommand{\hh}{H$_2$ }
\newcommand{\wave}{\,cm$^{-1}$ }
\newcommand{\degc}{$^\circ$C }
\newcommand{\fref}[1]{Fig.~\ref{#1}}


\title{Cold collisions in a molecular synchrotron}

\author{Aernout P.P. van der Poel$^1$}
\author{Peter C. Zieger$^{1,2}$}
\author{Sebastiaan Y.T. van de Meerakker$^3$}
\author{J\'{e}r\^{o}me Loreau$^4$}
\author{Ad van der Avoird$^3$}
\author{Hendrick L. Bethlem$^1$}
\affiliation{$^1$LaserLaB, Department of Physics and Astronomy, Vrije Universiteit, De Boelelaan 1081, 1081 HV Amsterdam, The Netherlands} 
\affiliation{$^2$Fritz-Haber-Institut der Max-Planck-Gesellschaft, Faradayweg 4-6, 14195 Berlin, Germany} 
\affiliation{$^3$Institute for Molecules and Materials, Radboud University, Heijendaalseweg 135, 6525 AJ Nijmegen, the Netherlands} 
\affiliation{$^4$Service de Chimie Quantique et Photophysique, Universit\'{e} Libre de Bruxelles (ULB) CP 160/09, 50 av. F.D. Roosevelt, 1050 Brussels, Belgium} 


\date{\today}

\begin{abstract}
We study collisions between neutral, deuterated ammonia molecules (ND$_3$) stored in a 50\,cm diameter synchrotron and argon atoms in co-propagating supersonic beams. The advantages of using a synchrotron in collision studies are twofold: (i) By storing ammonia molecules many round-trips, the sensitivity to collisions is greatly enhanced; (ii) The collision partners move in the same direction as the stored molecules, resulting in low collision energies. We tune the collision energy in three different ways: by varying the velocity of the stored ammonia packets, by varying the temperature of the pulsed valve that releases the argon atoms, and by varying the timing between the supersonic argon beam and the stored ammonia packets. These give consistent results. We determine the relative, total, integrated cross-section for $\mathrm{ND}_3+\mathrm{Ar}$ collisions in the energy range of 40--140\,cm$^{-1}$, with a resolution of 5--10\wave and an uncertainty of 7--15\%. Our measurements are in good agreement with theoretical scattering calculations.
\end{abstract}

\maketitle
The crossed molecular beam technique, pioneered by Dudley R. Herschbach and Yuan T. Lee, has yielded a detailed understanding of how molecules interact and react \cite{Lee:Science1987, Herschbach:ACIEE1987}. Until recently, these crossed molecular beam studies were limited by the velocities of the molecular beams to collision energies above a few 100\,cm$^{-1}$ (1\,cm$^{-1} \simeq 1.4$\,K). Over the last years, however, a number of ingenious methods \cite{Bell:MP2009, Narevicius:CR2012, Brouard:CSR2014} have been developed to study collisions in the cold regime. These advances are important for several reasons. Firstly, the temperatures of interstellar clouds are typically between 10--100\,K; collision data of simple molecules at low temperatures is thus highly relevant for understanding the chemistry in these clouds \cite{Roueff:CR2013}. Furthermore, quantum effects become important at low temperatures where few partial waves contribute and the de Broglie wavelength associated with the relative velocity becomes comparable to or larger than the intermolecular distances. Of particular interest are resonances of the collision cross-section as a function of collision energy \cite{Toennies:JCP1979, Chandler:JCP2010, Balakrishnan:JCP2000, Naulin:IRPC2014}. The position and shape of these resonances are very sensitive to the exact shape of the PES and thus serve as precise tests of our understanding of intermolecular forces.


The ability to control the velocity of molecules using time-varying electric and magnetic fields has allowed collision experiments at low collision energies \cite{Gilijamse:Science2006, Kirste:Science2012}. Recently, inelastic collisions between Stark-decelerated NO molecules and rare gas atoms have been studied with ion imaging techniques, revealing interesting quantum mechanical features such as Fraunhofer diffraction \cite{Onvlee:NC2016} and resonance fingerprints in the state-to-state differential cross-sections \cite{Vogels:Science2015}. 

Another method uses cryogenically cooled beams under a small (and variable) crossing angle. In recent experiments, scattering resonances were observed in collisions between O$_2$ or CO and H$_2$ molecules at energies between 5 and 30\,K \cite{Costes:CS2016}. Even lower temperatures can be obtained by using magnetic or electric guides to merge two molecular beams into a single beam \cite{Henson:Science2012,Osterwalder:EPJTI2015, Allmendinger:CPC2016}. This technique has been used to observe orbiting resonances in the Penning ionization reaction of argon and molecular hydrogen with metastable helium \cite{Henson:Science2012, Klein:NP2017}.

\begin{figure*}[t!]
	\centering
	\includegraphics[width=12.9cm]{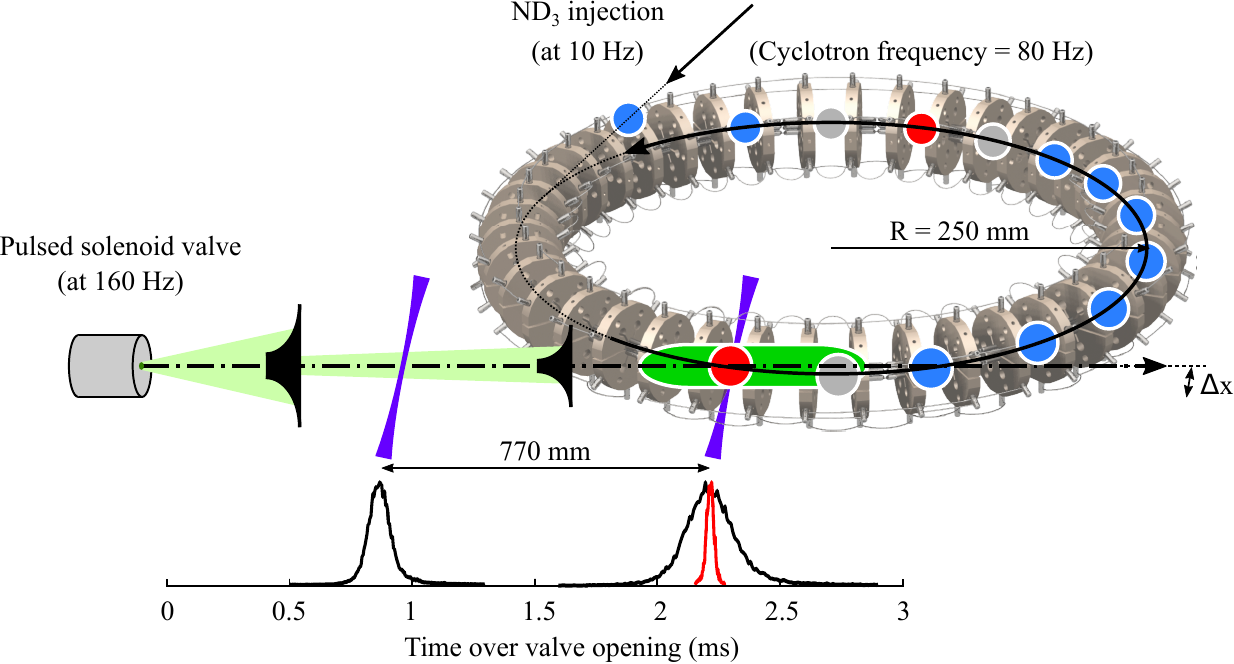}
	\caption{\label{fig:figure1}
		Schematic view of the synchrotron and beamline. Supersonic beams from a cooled Even-Lavie valve intersect the synchrotron a quarter round-trip down-stream from the injection point. The argon beam is displaced from the equilibrium orbit by a distance $\Delta x$. The packets of argon (shown in green) are timed such that the \emph{probe} packets (shown in red) encounter an argon packet on every round-trip, while \emph{reference} packets (shown in blue) provide a simultaneous measurement of background loss. Bottom: time-of-flight measurements of an argon beam at the two detection zones (black) and a packet of ammonia after 90 round-trips (red).}
\end{figure*}

In a different approach, trapped ions at millikelvin temperatures are monitored while slow, velocity selected, beams of molecules pass through the trap to study reactive ion-molecule collisions \cite{Willitsch:PRL2008, Chang:Science2013}. The ions are stored for a long time and their number can be accurately determined, which allows the study of reactions with rates as small as one per minute. In a similar fashion, collisions of slow beams of ammonia with magnetically trapped OH molecules were observed \cite{Sawyer:PCCP2011}.

Here, we study collisions between neutral ammonia molecules stored in a synchrotron and beams of argon atoms. Using a synchrotron for collision studies offers two advantages: Firstly, the collision partners move in the same direction as the stored molecules, resulting in a small relative velocity and thus a low collision energy. Secondly, the sensitivity to collisions is enhanced by storing the ammonia molecules for many round-trips. Our approach thus combines the low collision energies obtained in experiments that use merged molecular beams \cite{Henson:Science2012,Osterwalder:EPJTI2015, Allmendinger:CPC2016} with the high sensitivity of experiments that monitor trap loss \cite{Willitsch:PRL2008, Chang:Science2013, Sawyer:PCCP2011}.

The synchrotron used in our experiment is shown schematically in \fref{fig:figure1}. It consists of 40 electric hexapoles arranged in a circle with a diameter of 0.5\,m, to which voltages of up to $\pm 5$\,kV are applied. With these settings, we store packets of \nd molecules in the low-field seeking sublevel of the $J$=1, $K$=1 rovibrational ground state with velocities in the range of 100--150\,m/s. At any time during the experiment, 14 packets of ammonia molecules are held by the synchrotron. Each packet of this train is stored for up to 1.2\,s, while at a 10\,Hz rate new packets are being injected in front of the train and packets at the back of the train are being detected. In order to allow packets emerging from a Stark decelerator (not shown) to enter the ring, 4 hexapoles are temporarily switched off. In the detection zone, a quarter ring downstream from the injection point, molecules are ionized via the $\widetilde{\mathrm{B}}$-state using light around 317\,nm and pushed towards an ion detector by temporarily switching 2 hexapoles to the appropriate voltages. More details on the synchrotron and injection beamline can be found in Zieger \emph{et al.} \cite{Zieger:PRL2010, Zieger:PRA2013, Zieger:ZPC2013}. 

When the vacuum system is kept under vacuum for many weeks, the pressure reaches 5$\times$10$^{-9}$\,mbar. Under these conditions the 1/e-lifetime of the stored packets is 3.2\,s, determined equally by collisions with the background gas and black-body-radiation-induced transitions to non-trappable states \cite{Zieger:PRL2010}. In the current experiment the pressure is typically 2$\times$10$^{-8}$\,mbar, resulting in a lifetime of 1.0\,s. A typical time-of-flight profile of a packet that has been stored for 90 round-trips is shown in red in \fref{fig:figure1}. 

The stored packets of ammonia molecules are made to collide with beams of argon atoms released by a pulsed Even-Lavie valve \cite{Even:EPJTI2015}, which can be operated at temperatures between $-150$ and $+30^\circ$C. The pulsed valve is operated at 160\,Hz, two times the cyclotron frequency of the stored ammonia packets, such that every tenth \nd packet will encounter a fresh Ar packet every round-trip. We will refer to these packets as \emph{probe} packets. Packets that do not encounter the argon beam, referred to as \emph{reference} packets, provide a simultaneous measurement of background loss. Note that packets directly before or after the probe packets may have some overlap with the argon beam. Collisions between stored ammonia molecules and argon atoms will (almost always) cause ammonia molecules to be lost from the trap as the longitudinal and transverse trap depths ($\leq$1.1\,mK \cite{Zieger:ZPC2013}) are much smaller than the collision energies. The argon beam is displaced from the equilibrium orbit of the stored ammonia molecules by about 1.6\,mm such that it intersects the path of the ammonia molecules twice. This simplifies the analysis (\emph{vide infra}).

As shown in \fref{fig:figure1}, the relative intensities of the Ar beams are monitored at two positions: 480\,mm downstream from the valve and inside the detection zone of the synchrotron, 1250\,mm downstream from the valve. The argon atoms are detected by 3+1 Resonantly Enhanced Multi-Photon Ionization (REMPI) \emph{via} the $3s^23p^5(^2P_{1/2})4s$-state \cite{Minnhagen:JOSA1973} using a pulsed UV-laser running at 314\,nm. The arrival time distributions measured at the two detection zones are used to derive the longitudinal velocity distributions. The black curves in \fref{fig:figure1} show typical time of flight profiles measured in both detection zones for an argon beam at a valve temperature of $-150^\circ$C. 

\begin{figure}[b]
	\centering
	\includegraphics[width=8.6cm]{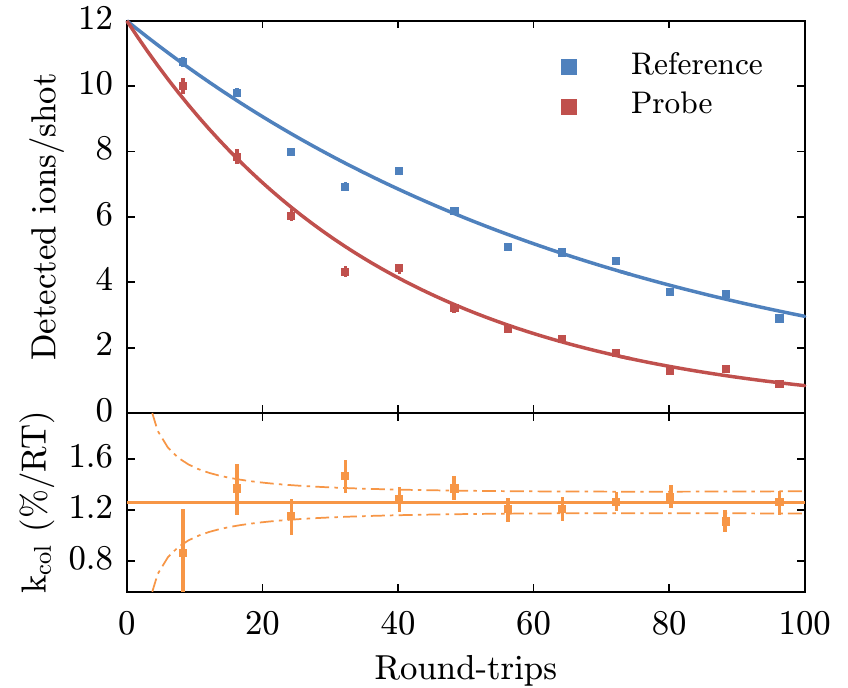}
	\caption{\label{fig:figure2}
		Upper frame: Number of detected \nd ions per shot as a function of the number of round-trips spent in the synchrotron, for both the probe (red) and the reference packets (blue). Each data point is an average over 3240 (reference) or 360 (probe) shots. The solid lines denote exponential fits. Lower frame: From ratios of the numbers of detected ions, the loss-rate due to collisions can be determined at each number of round-trips. The dash-dotted lines show the expected statistical uncertainty based on the number of detected ions per shot.}
\end{figure}

\fref{fig:figure2} shows the number of ammonia molecules in the probe beam (red squares) and reference beam (blue squares) as a function of storage time. In this particular experiment hydrogen molecules are used as collision partner. The solid lines show the result of fits to the data using the expression $n=n_{0}\,e^{\,k\cdot RT}$, with $n$ the number of detected ions per shot, $RT$ the number of round-trips the packets have made before being detected, and $k$ the loss-rate. For the reference packets we find a loss-rate of 1.40\% per round-trip, corresponding to a lifetime of 1.0\,s. For the probe packets, we find a loss-rate of 2.66\% per round-trip, which implies that collisions with the supersonic beam induce an additional loss-rate of 1.26\% per round-trip. 

The orange data points in the lower panel of \fref{fig:figure2} show the loss-rate due to collisions calculated from the number of ions detected in the probe and reference beams at specific round-trip numbers, using $k_{\mathrm{col}}= -\ln({n_{\mathrm{probe}}/n_\mathrm{ref}})/RT$. The error bars reflect the statistical spread of the ion signal. The uncertainty of $k_{\mathrm{col}}$ decreases dramatically during the first round-trips and reaches an optimum after $\sim$70 round-trips, at which point only 14\% of molecules in the probe beam remain. When the storage time is increased further, the uncertainty of the measured probe beam intensity becomes the limiting factor and the uncertainty of $k_{\mathrm{col}}$ increases. The orange dash-dotted lines show the expected statistical uncertainty, assuming the number of detected ions is governed by Poisson statistics, which is in good agreement with the experimental results. Note that the deviations from a perfect exponential decay of the probe and reference signals observed in the upper panel of \fref{fig:figure2} are absent in the extracted loss-rate shown in the lower panel. This is a crucial feature of our method: fluctuations due to, for instance, intensity and/or wavelength drifts of the laser, temperature variations of the valve, or collective oscillations of the packet inside the synchrotron are common to the reference and probe signals and are divided out. 

Of critical importance is the delay between the trigger of the valve that releases the argon atoms and the arrival time of the \nd probe packet in the detection zone. This delay determines whether the ammonia molecules collide with atoms located more in the leading or trailing end of the argon packet, or, in fact, whether they collide at all. Furthermore, as the flight time from the valve to the synchrotron is much larger than the opening time of the valve, there is a strong correlation between the position of the argon atoms and their velocity. Hence, the delay determines the velocity of the argon atoms that are encountered by the ammonia beam. Note that in our experiments, the relative velocities are such that ammonia molecules only see part (20--30\%) of the argon packet during the time they spend in the collision zone. 

\begin{figure*}[t]
	\centering
	\includegraphics[width=17.2cm]{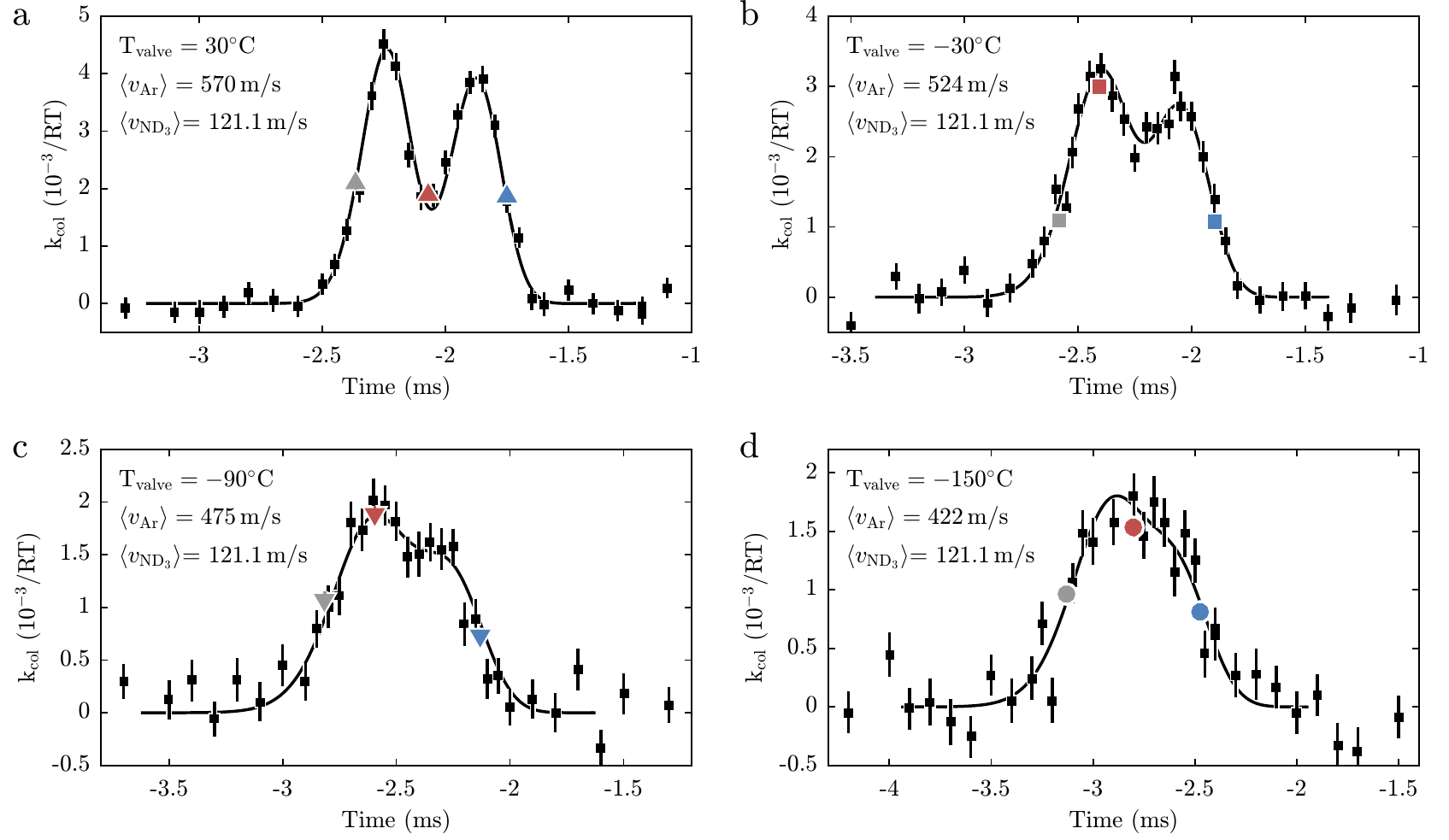}
	\caption{\label{fig:figure3}
		Loss-rate due to collisions with argon atoms versus valve timing for 4 different valve temperatures. The black squares consist of 2,400 shots each, the coloured points of 21,600 shots each. The lines depict results of a simulation of the experiment, individually scaled to fit the data (see main text).}
\end{figure*}

The loss-rates of \nd molecules (with velocities of 121.1\,m/s and 138.8\,m/s) due to collisions with argon atoms (with velocities around 570, 524, 475, and 422\,m/s) were measured as a function of the aforementioned delay. Each data point is the result of 2,400 shots--corresponding to a measurement time of 4 minutes. To be robust against possible drifts of the Ar beam density, the data were taken while toggling between the two ammonia speeds after every data point and picking the timings from a list in a random order. The results for an ammonia velocity of 121.1\,m/s are shown in panels a--d in \fref{fig:figure3}. Each measurement can be seen to feature two peaks, resulting from the fact that the argon beam intersects the synchrotron at two distinct locations. These peaks become less well resolved as the argon packets become slower and concomitantly longer.

Additional data were taken at three specific timings for both ammonia velocities. These data are shown as the colored symbols in \fref{fig:figure3}. Each of these is the result of 21,600 shots, corresponding to a measurement time of 36\,minutes per point. To detect and correct for possible drifts of the Ar beam density, we cycled nine times through the six different configurations. No significant drifts were detected.

The solid curves in \fref{fig:figure3} show results of a simulation of our experiment. The simulation uses as input (i) the longitudinal position and velocity distributions of the argon beams taken from the time-of-flight profiles at two locations, (ii) the shape and size of the cross-section of the argon beams taken from measurements where the height of the laser beam was scanned, from measurements where the position of the valve was scanned, and from the geometry of our beam machine, (iii) the horizontal displacement of the argon beam from the equilibrium orbit determined from the time difference between the maxima in the measured loss-rates, and (iv) the velocity and equilibrium radius of the synchronous molecule taken from simulations of the synchrotron \cite{Zieger:ZPC2013}. For each combination of valve temperature and ammonia velocity, the simulated curves are scaled vertically to fit the black data points in \fref{fig:figure3}. As seen from the figure, the simulations describe the measurements very well, which confirms that we have an excellent understanding of the experiment.

\begin{figure}[b]
	\centering
	\includegraphics[width=8.6cm]{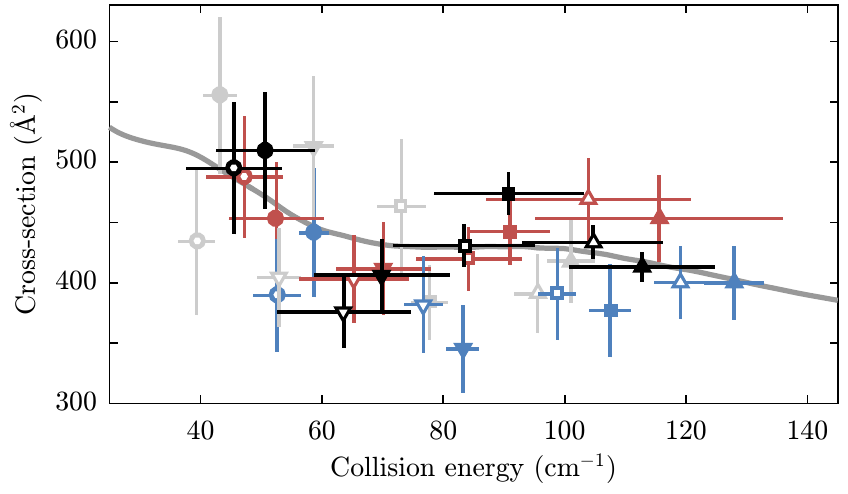}
	\caption{\label{fig:figure4}
		Total, integrated, $\mathrm{ND}_3+\mathrm{Ar}$ collision cross-section versus collision energy. The collision energy is varied in three different ways. The line shows a theoretical calculation  \cite{Loreau:JCP2015}, convoluted with a normal distribution with a standard deviation of 5\,cm$^{-1}$. The measurements are collectively fit to this calculation with a single global scaling factor.}
\end{figure}

The scaling factors obtained by fitting the simulations to the measurements are the products of the collision cross-sections and the absolute column densities of the argon beams at the corresponding valve temperatures. Hence, by dividing the scaling factors by the relative column densities of the argon beams, derived from time-of-flight measurements in the detection zone of the synchrotron, we obtain the relative total cross-sections. These are shown as the black data points in \fref{fig:figure4}. The blue, red, and gray points in \fref{fig:figure4} are found by scaling the simulations to the data points measured at the front, center, and back of the argon packet (the coloured points shown in \fref{fig:figure3}). The vertical error bars represent the statistical uncertainties of the measurements, ranging from 7 to 15\%, while the horizontal error bars represent the spread in collision energy, retrieved from the simulations, and range from 5 to 10\,cm$^{-1}$. The data taken in the fast part of the beam (the blue datapoints) have the best energy resolution. The data points taken by averaging over the entire velocity distribution have smaller uncertainties but their energy resolution is worse. The collision cross-sections determined from the different data sets agree with each other within their combined errors, although the cross-sections retrieved from colliding ammonia molecules with the fast part of the argon beam (blue data points) appear to be systematically smaller than the cross-sections retrieved from colliding ammonia molecules with the slow or center part of the argon beam.

The line depicted in \fref{fig:figure4} is the result of theoretical calculations described in \cite{Loreau:JCP2015}, convoluted with a normal distribution with a standard deviation of 5\,cm$^{-1}$. The measurements are collectively fit to this calculation with a single global scaling factor that represents the density of the argon beam at $\mathrm{T}=-150^\circ$C, which is found to be 4.1$\times$10$^{10}$\,cm$^{-3}$. Although the $\mathrm{ND}_3+\mathrm{Ar}$ collision cross-section in this energy range does not show spectacular features, the shallow minimum around 70\wave predicted by theory is reproduced in the experiment.

In conclusion, we have performed collision experiments between argon atoms in a supersonic beam and \nd molecules stored in a synchrotron. Our measurements demonstrate that storing molecules for many round-trips increases the sensitivity dramatically, while co-propagating beams allow low collision energies to be studied, hence providing a robust and general method to measure the total cross-section for low energy collisions. Our method has a number of additional features that make it attractive: (i) By comparing packets that are simultaneously stored in the synchrotron, the measurements are independent of the ammonia intensity and immune to variations of the background pressure in the synchrotron. (ii) As the probe packets interact with many argon packets, shot-to-shot fluctuations of the argon beam are averaged out. By toggling between different ammonia velocities and timings every 4\,minutes, slow drifts of the argon beam intensity are eliminated. The collision energy is currently limited by the large difference between the velocity of the stored molecules and the velocities in the supersonic beam. Lower collision energy could be reached by using molecules from cryostatically cooled beams as collision partner \cite{Hutzler:CR2012} and/or by using a larger synchrotron which would be able to store ammonia molecules at a higher velocity. Ideally, a synchrotron would be used that can store molecules directly from a supersonic beam without deceleration. Furthermore, one would like to store beams both clockwise and anticlockwise in the synchrotron, which makes it possible to perform calibration measurements at high energy. Note that if the velocities of the beams are more similar, the energy resolution will also be improved \cite{Shagam:JPCC2013}. Collision studies with paramagnetic atoms and molecules could be performed in a magnetic synchrotron \cite{VanderPoel:NJP2015}.

\begin{acknowledgments}
We thank Chris Eyles for help with setting up the synchrotron after being moved to Amsterdam. We thank Rob Kortekaas for technical support and Wim Ubachs and Gerard Meijer for helpful discussions and support. This research has been supported by the Netherlands Foundation for Fundamental Research of Matter (FOM) \emph{via} the program ``Broken mirrors and drifting constants''. S.Y.T.v.d.M. acknowledges support from the European Research Council under the European Union's Seventh Framework Programme (FP7/2007-2013) / ERC grant agreement 335646 MOLBIL.
\end{acknowledgments}

\end{document}